\documentclass[twocolumn,prd,nofootinbib]{revtex4}
\usepackage{amsmath}
\usepackage{amssymb}

\begin{document}

\title{Generalized Friedmann branes}
\author{L\'{a}szl\'{o} \'{A}. Gergely\thanks{
Laszlo.Gergely@port.ac.uk or gergely@physx.u-szeged.hu}}
\affiliation{Institute of Cosmology and Gravitation, University of Portsmouth, Portsmouth
PO1 2EG, UK and \\
Astronomical Observatory and Department of Experimental Physics, University
of Szeged, Szeged 6720, D\'{o}m t\'{e}r 9, Hungary}

\begin{abstract}
We prove that for a large class of generalized Randall-Sundrum II type
models the characterization of brane-gravity sector by the effective
Einstein equation, Codazzi equation and the twice-contracted Gauss equation
is equivalent with the bulk Einstein equation. We give the complete set of
equations in the generic case of non-$Z_{2}$-symmetric bulk and arbitrary
energy-momentum tensors both on the brane and in the bulk. Among these, the
effective Einstein equation contains a varying cosmological
\textquotedblleft constant\textquotedblright\ and two new source terms. The
first of these represents the deviation from $Z_{2}$ symmetry, while the
second arises from the bulk energy-momentum tensor. We apply the formalism
for the case of perfect fluid on a Friedmann brane embedded in a generic
bulk. The generalized Friedmann and Raychaudhuri equations are given in a
form independent of both the embedding and the bulk matter. They contain two
new functions obeying a first order differential system, both depending on
the bulk matter and the embedding. Then we focus on Friedmann branes
separating two non-identical (inner or outer) regions of Reissner-Nordstr%
\"{o}m-Anti de Sitter bulk space-times, generalizing previous non-$Z_{2}$%
-symmetric treatments. Finally the analysis is repeated for the Vaidya-Anti
de Sitter bulk space-time, allowing for both ingoing and outgoing radiation
in each region.
\end{abstract}

\date{\today}
\startpage{1}
\endpage{}
\maketitle

\section{Introduction}

Since the pioneering idea of Randall and Sundrum \cite{RS2} of enriching the
four dimensional space-time with one noncompact spatial dimension, cosmology
has advanced towards new types of extensions. Generalized Randall-Sundrum II
type models have in common a five dimensional space-time (bulk), governed by
the Einstein equations, and a four dimensional brane, representing our
physical world, on which ordinary matter fields are confined. At low
energies gravity is also localized at the brane \cite{RS2}, however this
feature does not always hold \cite{Dadhich}. Generalizations of the original
Randall-Sundrum scenario are various and multiple, all allowing for matter
with cosmological symmetry on the brane (Friedmann branes) \cite{Maartens}, 
\cite{BDEL} (in this case the bulk is Schwarzschild-Anti de Sitter
space-time \cite{MSM}, \cite{BCG}). The assumption of $Z_{2}$ symmetric
embedding was also lifted in a series of papers \cite{Kraus}-\cite{Stoica},
and non-empty bulks have also been considered, with physically reasonable
matter content, like null dust \cite{ChKN}, \cite{LSR}, which can be
interpreted as the high frequency (geometrical optics) approximation of
unpolarized radiation (even gravitational), whenever the wavelength of the
radiation is negligible compared to the curvature radius of the background.
In the present paper we present a formalism generic enough to allow for all
such types of extensions. Models allowing a \textit{dilatonic} type scalar
field in the bulk were also discussed \cite{MW}, but will not be dealt with
in the context of this paper, neither will the possibility of having
different coupling constants on the two sides of the brane \cite{Stoica}.
Further generalization of our formalism however, is straightforward.

In Section 2 we present the decomposition of the Einstein tensor in an
arbitrary $\left( d+1\right) $ dimensional space-time with respect to some
generic (timelike or spacelike foliation). We carefully monitor the
relationship of the tensor, vector and scalar projections of the Einstein
equation with the system of effective Einstein and Codazzi equations, widely
employed in brane-world scenarios. We show that the latter system should be
supplemented by the twice contracted Gauss equation in order to assure the
full equivalence. (In this context we mention a recent analysis \cite%
{AndersonTavakol}, which also underlines the unsatisfactory feature of
\textquotedblleft truncating\textquotedblright\ the system of bulk Einstein
equations to brane equations.)

Beginning with Section 3 we have in mind the brane-world scenario. By use of
the Lanczos-Sen-Darmois-Israel junction conditions \cite{Lanczos}-\cite%
{Israel} we derive the generalized effective Einstein equation in a form
closely resembling previous works \cite{SMS}: 
\begin{equation}
G_{ab}=-\Lambda g_{ab}+\kappa ^{2}T_{ab}+\widetilde{\kappa }^{4}S_{ab}-%
\overline{\mathcal{E}}_{ab}+\overline{L}_{ab}^{TF}+\overline{\mathcal{P}}%
_{ab}\ .  \label{modEgen}
\end{equation}%
Among the source terms on the rhs we find the brane energy-momentum tensor $%
T_{ab}$, the term $S_{ab}$ quadratic in the energy-momentum tensor (relevant
at high energies) and $\overline{\mathcal{E}}_{ab}$, the electric part of
bulk Weyl tensor. Our generic treatment \textit{does not} require the $Z_{2}$
symmetry of the bulk across the brane and this leads to three important
modifications. First $\overline{\mathcal{E}}_{ab}$ represents an average
taken over the two sides of the brane. Second, a new source term $\overline{L%
}_{ab}^{TF}$ appears. Third, there is a contribution included in $\Lambda $,
which transforms the\ cosmological \textquotedblleft
constant\textquotedblright\ into a function. Bulk energy-momentum is also
allowed, resulting in the $\overline{\mathcal{P}}_{ab}$ source term and a
second non-constant contribution to $\Lambda $. When allowing bulk matter,
we have in mind reasonable sources, like null dust or multi-component null
dust, which can model for example the cross-flow of gravitational radiation
escaping the brane and Hawking radiation leaving the bulk black hole(s). At
the end of the section we give the generic form of the brane Bianchi
identities.

The most interesting applications of the developed formalism would be for
branes containing black holes and for branes containing perfect fluid and
obeying cosmological symmetries. Anisotropic cosmological brane-world models
can also be considered \cite{Frolov}, \cite{CMMS}.

Among these in Section 4 we discuss the case of maximally symmetric branes
with perfect fluid. By employing the brane Bianchi identities, we derive the
generalized Raychaudhuri and Friedmann equations \textit{in a form
insensitive to both the choice of the bulk matter and of the left and right
embeddings of the brane in the bulk}. The price to pay is a cosmological
function instead of a constant, and that the source term usually quoted as
dark radiation acquires a much broader interpretation.

An algorithm to solve in a hierarchical way the relevant system of equations
for a bulk containing a Friedmann brane with a given perfect fluid is
presented in the Appendix. The equations refer to $d=4$. The algorithm is
suited for the cases when no a priori choice of the bulk is performed,
instead the matter content of the bulk and the details of the embedding are
specified. There are constraints on both of these choices, as detailed in
the Appendix.

Section 5 deals with Friedmann branes embedded in the Reissner-Nordstr\"{o}%
m-Anti de Sitter bulk. In the case of a cosmological bulk with maximmally
symmetric spatial 3 sections (case without the charge), with the exception
of the static case, where exotic solutions are equally possible \cite%
{Einbrane}, a generalized Birkhoff theorem holds \cite{BCG}, which states
that such a bulk is the 5d Schwarzschild-anti de Sitter space-time. We
develop a formalism which is suitable for matching inner and outer regions
of the Reissner-Nordstr\"{o}m-Anti de Sitter space-time, thus allowing for
two, one or no charged black hole. We give the energy-momentum tensor
leading to the solution already employed in \cite{BVisser} in the study of
the $Z_{2}$ symmetric embedding. Then by straightforward algebra we find the
generalized Friedmann and Raychaudhuri equations. These were checked to
reproduce all previous non-$Z_{2}$ symmetric results derived in the
particular case of pure cosmological bulk: different black hole masses on
the two sides on the brane \cite{Kraus}, \cite{Ida}, \cite{Davis}, zero
black hole mass and different cosmological constants on the two sides \cite%
{Deruelle}, \cite{Perkins} and allowing for both types of generalizations 
\cite{Stoica}, \cite{BCG}, \cite{Carter}.

In Section 6 we study a generic \textit{asymptotically} anti de Sitter bulk
compatible with type II fluid. Such a bulk is a generalzation of the charged
Vaidya solution in the presence of a cosmological constant. In \cite%
{Kaminaga} the charged Vaidya solution was employed to model evaporating
charged black holes. Here first we derive the 5 dimensional solution, which
agrees with \cite{ChKN}, generated by null dust and an electromagnetic field
on a cosmological background. The details of these source terms were not
given in the literature before. We then follow the generic prescription
described in the Appendix and we write down the generalized Friedmann and
Raychaudhuri equations for this case. The generic results derived here are
also new. In the particular case of $Z_{2}$ symmetric brane embeddings, the
Friedmann equation for Vaidya-anti de Sitter bulk \cite{ChKN} is recovered.

Throughout the paper a tilde distinguishes the quantities defined on the $%
\left( d+1\right) $ dimensional space-time. The only exception is the normal 
$n$ to the leaves of the foliation. Its norm is $n^{c}n_{c}=\epsilon =\pm 1$
($\epsilon =1$ stands for timelike and $\epsilon =-1$ for spacelike
foliations). Latin indices represent abstract indices running from $0$ to $d$%
. Vector fields in Lie-derivatives are represented by boldface characters.
For example $\widetilde{\mathcal{L}}_{\widetilde{\mathbf{V}}}T$ denotes the
Lie derivative along the integral lines of the vector field $\widetilde{V}%
^{a}$. From section to section, as the paper converges toward its conclusion
and the results derived apply to more specific situations, the degree of
generality decreases accordingly. Up to Section 3.b everything holds for
arbitrary $\epsilon $. Up to and including Section 4, the results are
dimension-independent, $d=4$ being imposed only in the application described
in Sections 5 and 6 and in the Appendix. In Section 6, $\epsilon $
distinguishes between the outgoung or ingoing character of the radiation.

\section{The equivalence of two $\left( d+1\right) $ dimensional
decompositions}

The $\left( d+1\right) $-metric $\widetilde{g}_{ab}$ induces a metric $%
g_{ab} $ on the leaves, 
\begin{equation}
\widetilde{g}_{ab}=g_{ab}+\epsilon n_{a}n_{b}\ .  \label{tildeg0}
\end{equation}%
By introducing the projectors $%
g_{c_{1}...c_{r}b_{1}...b_{s}}^{a_{1}...a_{r}d_{1}...d_{s}}=g_{c_{1}}^{a_{1}}{}...g_{c_{r}}^{a_{r}}{}g_{b_{1}}^{d_{1}}{}...g_{b_{s}}^{d_{s}}{} 
$, one can define the projected covariant derivative and the projected
Lie-derivative of any tensor $\widetilde{T}_{b_{1}...b_{s}}^{a_{1}...a_{r}}$
as 
\begin{align}
\nabla _{a}\widetilde{T}_{b_{1}...b_{s}}^{a_{1}...a_{r}}&
=g_{ac_{1}...c_{r}b_{1}...b_{s}}^{ca_{1}...a_{r}d_{1}...d_{s}}\widetilde{%
\nabla }_{c}\widetilde{T}_{d_{1}...d_{s}}^{c_{1}...c_{r}}\ ,  \label{cov} \\
\mathcal{L}_{\widetilde{\mathbf{V}}}\widetilde{T}%
_{b_{1}...b_{s}}^{a_{1}...a_{r}}&
=g_{c_{1}...c_{r}b_{1}...b_{s}}^{a_{1}...a_{r}d_{1}...d_{s}}\widetilde{%
\mathcal{L}}_{\widetilde{\mathbf{V}}}\widetilde{T}%
_{d_{1}...d_{s}}^{c_{1}...c_{r}}\ .  \label{Lie}
\end{align}%
If both the tensor $\widetilde{T}_{b_{1}...b_{s}}^{a_{1}...a_{r}}$ and the
vector $\widetilde{V}^{a}$ are defined on the leaves, the above equations
are the $g_{ab}$-compatible covariant derivative and Lie-derivative on the
lower-dimensional space, respectively. If $\widetilde{V}^{a}$ however is
transverse to the leaves, the projected Lie-derivative describes transverse
evolution. The embedding of the leaves in the $\left( d+1\right) $
dimensional space-time is characterized by the extrinsic curvature $%
K_{ab}=\nabla _{a}n_{b}.$ Its trace will be denoted by $K$. It is immediate
to see that $K_{ab}$ is symmetric by noting that $n_{b}=\beta \widetilde{%
\nabla }_{b}\chi $ ($\beta $ is an arbitrary function; as the condition $%
\chi $=const. defines the leaves, $\chi $ is time for $\epsilon =-1$ and any
coordinate transverse to the leaves for $\epsilon =1$). The extrinsic
curvature obeys 
\begin{equation}
2K_{ab}=\mathcal{L}_{\mathbf{n}}g_{ab}\ .  \label{KLie}
\end{equation}%
The congruence $n^{a}$ has its own curvature \cite{Schouten} $\alpha
^{b}=n^{c}\widetilde{\nabla }_{c}n^{b}=g_{d}^{b}\alpha ^{d}$. For spacelike
foliations this is the nongravitational acceleration of observers with
velocity $n^{a}$. With this we find 
\begin{equation}
\widetilde{\nabla }_{a}n_{b}=K_{ab}+\epsilon n_{a}\alpha _{b}\ .
\label{normderivdecomp}
\end{equation}%
For notational convenience we also introduce the tensors 
\begin{align}
E_{ab}& =K_{ac}K_{b}^{c}-\mathcal{L}_{\mathbf{n}}K_{ab}+\nabla _{b}\alpha
_{a}-\epsilon \alpha _{b}\alpha _{a}\ ,  \label{E} \\
F_{ab}& =KK_{ab}-K_{ac}K_{b}^{c}\ ,  \label{F}
\end{align}%
together with their traces, $E$ and $F$. Note that $E_{ab}$ carries the
information about the transverse evolution of $K_{ab}$.

The $\left( d+1\right) $-dimensional Einstein tensor is equivalent with the
following set of projections 
\begin{subequations}
\label{Eindecomp}
\begin{align}
g_{a}^{c}{}g_{b}^{d}{}\widetilde{G}_{cd}& =G_{ab}-\epsilon \left[
F_{ab}-E_{ab}-\frac{1}{2}g_{ab}\left( F-2E\right) \right] \ ,  \label{tensor}
\\
g_{a}^{c}{}n^{d}{}\widetilde{G}_{cd}& =g_{a}^{c}n^{d}\widetilde{R}%
_{cd}=\nabla _{c}K_{a}^{c}-\nabla _{a}K\ ,  \label{vector} \\
2n^{a}n^{b}\widetilde{G}_{ab}& =-\epsilon R+F\ .  \label{scalar}
\end{align}%
These equations have the following meaning, provided $\widetilde{G}_{ab}$ is
determined by a $\left( d+1\right) $-dimensional Einstein equation. The
tensor equation (\ref{tensor}) determines (through $E_{ab}$) the evolution
of $K_{ab}$ normal to the foliation. Together with (\ref{KLie}) they give
the off-leave evolutions of the variables $\left( g_{ab,}K_{ab}\right) $
defined on the leaves. The vector equation (\ref{vector}) is the Codazzi
equation and represents a constraint on these variables. Similarly does the
scalar equation (\ref{scalar}). For spacelike foliations the vector and
scalar equations are the diffeomorphism and Hamiltonian constraints,
respectively. These\ \textquotedblleft instantaneous
constraints\textquotedblright\ become dynamical for timelike foliations and
the evolution equations form an elliptic, rather than hyperbolic system.

In what follows we would like to set up an equivalent set of equations, most
commonly employed in brane-world scenarios, e.g. suitable for timelike
foliations. We would like to keep, however a strict account of the sets of
equations which are equivalent to each other in the two pictures. For this
purpose first we decompose the tensor equation (\ref{tensor}) into its trace 
\end{subequations}
\begin{equation}
2\epsilon g^{ab}{}\widetilde{G}_{ab}=\left( d-2\right) \left( -\epsilon
R+F\right) -2\left( d-1\right) E\ ,  \label{trace}
\end{equation}%
and trace-free parts 
\begin{equation}
-\epsilon \left( g_{a}^{c}{}g_{b}^{d}{}\widetilde{G}_{cd}\right)
^{TF}=\left( -\epsilon R_{ab}+F_{ab}-E_{ab}\right) ^{TF}\ ,
\label{traceless}
\end{equation}%
where $^{TF}$ denotes tracefree, e.g. $f_{ab}^{TF}=f_{ab}-fg_{ab}/d$ for any
tensor $f_{ab}$ defined on the leaves.

The trace equation (\ref{trace}), properly combined with the scalar equation
(\ref{scalar}) gives the twice contracted Gauss equation: 
\begin{equation}
-\epsilon \widetilde{R}=-\epsilon R+F-2E\ .  \label{Gauss2}
\end{equation}%
Eliminating $R$ from Eqs. (\ref{scalar}) and (\ref{Gauss2}) gives $E$ solely
in terms of bulk tensors: 
\begin{equation}
E=n^{a}n^{b}\widetilde{G}_{ab}+\frac{\epsilon }{2}\widetilde{R}\ .
\label{Edef}
\end{equation}%
Concerning the tracefree part $E_{ab}^{TF}$, it is commonly expressed in
terms of the Weyl tensor (the purely radiative contribution to gravity) 
\begin{eqnarray}
\widetilde{C}_{abcd} &=&\widetilde{R}_{abcd}+\frac{2}{d\left( d-1\right) }%
\widetilde{g}_{a[c}\widetilde{g}_{d]b}\widetilde{R}  \notag \\
&&-\frac{2}{d-1}\left( \widetilde{g}_{a[c}\widetilde{R}_{d]b}-\widetilde{g}%
_{b[c}\widetilde{R}_{d]a}\right) \ .  \label{Weyl5}
\end{eqnarray}%
Its \textquotedblleft electric\textquotedblright\ part with respect to $%
n^{a} $ is defined as 
\begin{eqnarray}
\mathcal{E}_{ac} &=&\widetilde{C}_{abcd}n^{b}n^{d}\
=g_{a}^{i}n^{j}g_{c}^{k}n^{l}\widetilde{R}_{ijkl}+\frac{\epsilon \widetilde{R%
}}{d\left( d-1\right) }g_{ac}  \notag \\
&&-\frac{1}{d-1}\left( \epsilon g_{a}^{i}g_{c}^{k}\widetilde{R}%
_{ik}+g_{ac}n^{i}n^{k}\widetilde{R}_{ik}\right) \ .
\end{eqnarray}%
Inserting the projections 
\begin{align}
g_{a}^{i}n^{j}g_{c}^{k}n^{l}\widetilde{R}_{ijkl}& =E_{ac}\ ,  \notag \\
-\epsilon g_{a}^{i}g_{c}^{k}\widetilde{R}_{ik}& =-\epsilon
R_{ac}+F_{ac}-E_{ac}\ ,  \notag \\
n^{i}n^{k}\widetilde{R}_{ik}& =-K_{bd}K^{bd}-E\ ,  \label{proj}
\end{align}%
we find 
\begin{equation}
\left( d-1\right) \mathcal{E}_{ab}=\left[ -\epsilon R_{ab}+F_{ab}+\left(
d-2\right) E_{ab}\right] ^{TF}\ .  \label{eps}
\end{equation}%
Eliminating $R_{ab}$ from Eqs. (\ref{traceless}) and (\ref{eps}) leads to 
\begin{equation}
\mathcal{E}_{ab}+\epsilon \frac{1}{d-1}\left( g_{a}^{c}{}g_{b}^{d}{}%
\widetilde{G}_{cd}\right) ^{TF}=E_{ab}^{TF}\ .  \label{eps1}
\end{equation}%
In what follows, this equation containing the tracefree part of the
off-leave evolution $\mathcal{L}_{\mathbf{n}}K_{ab}$ will be regarded as the
definition of $\mathcal{E}_{ab}$. Eliminating the off-leave derivative term
from Eqs. (\ref{traceless}) and (\ref{eps}) results in 
\begin{equation}
\mathcal{E}_{ab}-\epsilon \frac{d-2}{d-1}\left( g_{a}^{c}{}g_{b}^{d}{}%
\widetilde{G}_{cd}\right) ^{TF}=\left( -\epsilon R_{ab}+F_{ab}\right) ^{TF}\
.  \label{traceless1}
\end{equation}%
Combining this trace-free equation with the scalar equation (\ref{scalar})
we obtain the effective Einstein equation on the leaves: 
\begin{eqnarray}
G_{ab} &=&\frac{d-2}{d-1}\left( g_{a}^{c}{}g_{b}^{d}{}\widetilde{G}%
_{cd}\right) ^{TF}+\frac{d-2}{d}g_{ab}\epsilon n^{c}n^{d}\widetilde{G}_{cd} 
\notag \\
&&+\epsilon \left( F_{ab}-\frac{g_{ab}}{2}F\right) -\epsilon \mathcal{E}%
_{ab}\ .  \label{modE}
\end{eqnarray}%
Note that the trace of the effective Einstein equation (\ref{modE}) and the
scalar equation (\ref{scalar}) coincide by construction. Therefore the
second scalar equation is given by the trace of the original tensor equation
(\ref{tensor}) which, as we have seen, is equivalent (modulo the trace of
the effective Einstein equation) to either the twice-contracted Gauss
equation (\ref{Gauss2}) or to Eq. (\ref{Edef}).

For spacelike foliations the usual way to think of the above system of
equations is to choose variables $g_{ab}$ and $K_{ab}$ satisfying the
constraints (\ref{vector}) and (\ref{scalar}) on the leaves and let them
evolve via Eqs. (\ref{KLie}) and (\ref{tensor}). When the foliation is
timelike, another viewpoint is common. In the brane-world scenario the
central role is played by the effective Einstein equation (\ref{modE}), in
which the bulk matter (via the bulk Einstein equation), the extrinsic
curvature of the brane (the $F$-terms) and the electric part of the bulk
Weyl tensor are all considered sources for the brane gravity sector. While
the extrinsic curvature of the brane is determined by brane matter and brane
tension through the junction mechanism, and (in the $Z_{2}$ symmetric
cosmological bulk) the longitudinal part of $\mathcal{E}_{ab}$ is fixed by
the vector equation \cite{SMS}, nothing constraints the behavior of the
transverse part of $\mathcal{E}_{ab}$, which remains arbitrary from a brane
point of view. This feature is the source of several difficulties,
frequently formulated as the lack of a \textit{temporal} evolution equation
for $\mathcal{E}_{ab}$.

The \textit{off-brane} evolution of $\mathcal{E}_{ab}$ was deduced from the
bulk Bianchi identities in Ref. \cite{SMS} (in the case of a $Z_{2}$
symmetric cosmological bulk). It also follows from the above equations. As $%
\mathcal{E}_{ab}$ is completely determined by the bulk matter, the induced
metric and the extrinsic curvature via Eq. (\ref{traceless1}), the evolution
of $\mathcal{E}_{ab}$ follows from the metric evolution (\ref{KLie}) and the
evolution of the extrinsic curvature. Let us recall that the latter was
given by the tensor equation (\ref{tensor}). Should one choose a brane-world
viewpoint, the situation is different: the effective Einstein equation (\ref%
{modE}) together with Eq. (\ref{eps1}) gives only the traceless part of $%
\mathcal{L}_{\mathbf{n}}K_{ab}.$ The complementary equation is either the
twice contracted Gauss equation (\ref{Gauss2}) or Eq. (\ref{Edef}), which
both contain $g^{ab}\mathcal{L}_{\mathbf{n}}K_{ab}$. With this equation, the
bulk Bianchi identities emerge as a consequence.

It is clear now that in the brane-world scenario the effective Einstein
equation and the Codazzi equations do not provide a complete
characterization of gravity, but they should be supplemented by the
twice-contracted Gauss equation (or the expression (\ref{Edef}) for $E$).
This set of equations is equivalent with the Einstein equation in the $%
\left( d+1\right) $ dimensional space-time.

\section{The effective Einstein equation for non-$Z_{2}$-symmetric bulk}

\subsection{The junction conditions}

Both in general relativity and in the brane world scenarios the possibility
of a distributional matter source on a hypersurface is of interest. Such a
hypersurface divides the space-time into two distinct regions. In both of
these regions one of the systems (\ref{Eindecomp}) or (\ref{vector}), (\ref%
{Edef}) and (\ref{modE}) should be imposed separately. Quantities defined on
these domains will be distinguished by $+$ or $-$ symbols. The passage from
one zone to the other is described in a coordinate-independent manner by the
junction conditions \cite{Israel} (see also \cite{BCG}). These conditions
include the continuity of the induced metric across the hypersurface, $%
g_{ab}^{+}=g_{ab}^{-}$, and the Lanczos equation \cite{Lanczos}, a condition
on the jump of the extrinsic curvature. It is straightforward to deduce the
latter equation from the equations derived in the preceding section.

From (\ref{E}), (\ref{F}) and the second relation (\ref{proj}) we find 
\begin{align}
\mathcal{L}_{\mathbf{n}}K_{ab}& =-\epsilon g_{a}^{i}g_{b}^{k}\widetilde{R}%
_{ik}+Z_{ab}\ ,  \label{LieK} \\
Z_{ab}& =\epsilon R_{ab}+2K_{ac}K_{b}^{c}-KK_{ab}+\nabla _{b}\alpha
_{a}-\epsilon \alpha _{b}\alpha _{a}\ .
\end{align}%
The Einstein equation gives 
\begin{equation}
g_{a}^{i}g_{b}^{k}\widetilde{R}_{ik}=\widetilde{\kappa }^{2}\left(
g_{a}^{i}g_{b}^{k}\widetilde{T}_{ik}-\frac{1}{d-1}g_{ab}\widetilde{T}\right)
\ .
\end{equation}%
If $l$ is the coordinate adapted to the normal, $\mathbf{n=}\partial
/\partial l$, the energy-momentum tensor can be written as $\widetilde{T}%
_{ik}=\widetilde{\Pi }_{ik}+\tau _{ik}\delta \left( l\right) $, with $%
\widetilde{\Pi }_{ik}$ the regular part and $\tau _{ik}$ the distributional
part on the layer, obeying $\tau _{ik}n^{i}=0$. Thus (\ref{LieK}) becomes 
\begin{align}
\frac{\partial }{\partial l}K_{ab}& =-\epsilon \widetilde{\kappa }^{2}\left(
\tau _{ab}-\frac{1}{d-1}g_{ab}\tau \right) \delta \left( l\right)
+W_{ab}+Z_{ab}\ , \\
W_{ab}& =-\epsilon \widetilde{\kappa }^{2}\left( g_{a}^{i}g_{b}^{k}%
\widetilde{\Pi }_{ik}-\frac{1}{d-1}g_{ab}\widetilde{\Pi }\right) \ .
\end{align}%
As both $Z_{ab}$ and $W_{ab}$ are finite, integration across the layer on an
infinitesimal integration range gives the Lanczos equation: 
\begin{equation}
\Delta K_{ab}=-\epsilon \widetilde{\kappa }^{2}\left( \tau _{ab}-\frac{1}{d-1%
}g_{ab}\tau \right) \ ,  \label{Lanczos}
\end{equation}%
or equivalently 
\begin{equation}
-\epsilon \widetilde{\kappa }^{2}\tau _{ab}=\Delta K_{ab}-g_{ab}\Delta K\ .
\label{Lanczos1}
\end{equation}%
Here we have introduced the notation$\ \Delta f_{ab}=f_{ab}^{+}-f_{ab}^{-}$
for the jump of any tensor $f_{ab}$ and $\Delta f$ for its trace. (By
construction, $+$ means the region towards which $n$ is pointing. We
emphasize, that the Lanczos equation is not affected by the choice of the
orientation of the normal $n$, because the change in the orientation implies
that both the $+$ and $-$ regions and the sign of the extrinsic curvature
are reversed.) We also introduce the mean value $\overline{f}_{ab}=\left(
f_{ab}^{+}+f_{ab}^{-}\right) /2$. Obviously $\Delta g_{ab}=0$ and $\overline{%
g}_{ab}=g_{ab}$. Straightforward algebra then shows 
\begin{align}
\overline{F}_{ab}& =\overline{K}_{ab}\overline{K}-\overline{K}_{ac}\overline{%
K}_{b}^{c}+\delta F_{ab}\ ,  \notag \\
\overline{F}& =\overline{K}^{2}-\overline{K}_{ab}\overline{K}^{ab}+\delta F\
,  \notag \\
\Delta F_{ab}& =-\epsilon \widetilde{\kappa }^{2}\Biggl[\overline{K}\left(
\tau _{ab}-\frac{1}{d-1}g_{ab}\tau \right)  \notag \\
& +\overline{K}_{ab}\frac{\tau }{d-1}-2\overline{K}_{c(a}\tau _{b)}^{c}%
\Biggr]\ ,  \notag \\
\Delta F& =2\epsilon \widetilde{\kappa }^{2}\overline{K}_{ab}\tau ^{ab}\ ,
\label{Fmeanjump}
\end{align}%
where we have denoted by 
\begin{equation}
\delta F_{ab}=-\frac{\widetilde{\kappa }^{4}}{4}\left( \tau _{ac}\tau
_{b}^{c}-\frac{1}{d-1}\tau \tau _{ab}\right) \ 
\end{equation}%
the contribution which distinguishes the functional form of $\overline{F}%
_{ab}$ from the one of $F_{ab}$.

Let us now consider a region of space-time of a finite thickness $2\eta $,
which contains this temporal hypersurface. The set of equations (\ref{vector}%
) and (\ref{modE}) holds in any of the two regions even in the limit $\eta
\rightarrow 0.$ Their sum and difference give: 
\begin{subequations}
\label{system}
\begin{align}
\widetilde{\kappa }^{2}\overline{\left( g_{a}^{c}{}n^{d}\widetilde{\Pi }%
_{cd}\right) }& =\nabla _{c}\overline{K}_{a}^{c}-\nabla _{a}\overline{K}\ ,
\label{vp} \\
\Delta \left( g_{a}^{c}{}n^{d}{}\widetilde{\Pi }_{cd}\right) & =-\epsilon
\nabla _{c}\tau _{a}^{c}\ ,  \label{vm} \\
2\widetilde{\kappa }^{2}\overline{\left( n^{a}n^{b}\widetilde{\Pi }%
_{ab}\right) }& =-\epsilon R+\overline{F}\ ,  \label{trp} \\
\Delta \left( n^{a}n^{b}\widetilde{\Pi }_{ab}\right) & =\epsilon \overline{K}%
_{ab}\tau ^{ab}\ ,  \label{trm} \\
\widetilde{\kappa }^{2}\frac{d-2}{d-1}\overline{\left( g_{a}^{c}{}g_{b}^{d}{}%
\widetilde{\Pi }_{cd}\right) ^{TF}}& =R_{ab}^{TF}-\epsilon \overline{F}%
_{ab}^{TF}+\epsilon \overline{\mathcal{E}}_{ab}\ ,  \label{trlessp} \\
\widetilde{\kappa }^{2}\frac{d-2}{d-1}\Delta \left( g_{a}^{c}{}g_{b}^{d}{}%
\widetilde{\Pi }_{cd}\right) ^{TF}& =-\epsilon \Delta F_{ab}^{TF}+\epsilon
\Delta \mathcal{E}_{ab}\ .  \label{trlessm}
\end{align}%
The last four equations are the trace and trace-free parts of the sum and
difference of the effective Einstein equations in the two regions,
respectively. From among them the last two equations define the mean value
and the jump of $\mathcal{E}_{ab}$ (the trace-free part of $E_{ab}$). Let us
recall that the trace $E$ is also determined in both regions by Eq. (\ref%
{Edef}), which in terms of the bulk energy-momentum tensor reads: 
\end{subequations}
\begin{equation}
E^{\pm }=\widetilde{\kappa }^{2}\left( n^{a}n^{b}\widetilde{\Pi }_{ab}^{\pm
}-\frac{\epsilon }{d-1}\widetilde{\Pi }^{\pm }\right) \ .  \label{Edefpm}
\end{equation}%
As it will be employed in the next subsection, we also give the undecomposed
form of the equation obtained by the sum of the effective Einstein equations
on each side (equivalent to Eqs. (\ref{trp}) and (\ref{trlessp})): 
\begin{eqnarray}
G_{ab} &=&\widetilde{\kappa }^{2}\left[ \frac{d-2}{d-1}\overline{\left(
g_{a}^{c}{}g_{b}^{d}{}\widetilde{\Pi }_{cd}\right) ^{TF}}+\epsilon \frac{d-2%
}{d}g_{ab}\overline{\left( n^{c}n^{d}\widetilde{\Pi }_{cd}\right) }\right] 
\notag \\
&&+\epsilon \left( \overline{F}_{ab}-\frac{g_{ab}}{2}\overline{F}-\overline{%
\mathcal{E}}_{ab}\right) \ .  \label{modEp}
\end{eqnarray}%
\qquad

\subsection{The effective Einstein equation}

From now on we specialize to the brane-world scenarios, where a $\left(
d-1\right) $-dimensional distributional source evolves in time and in
consequence the hypersurface is temporal. Thus we apply the above formulae
for $\epsilon =1.$ For the generic brane energy-momentum tensor $\tau
_{ab}=-\lambda g_{ab}+T_{ab}$ (where $\lambda $ is the brane tension and $%
T_{ab}$ represents ordinary matter on the brane) we have: 
\begin{eqnarray}
\delta F_{ab}-\frac{g_{ab}}{2}\delta F &=&\widetilde{\kappa }^{4}\Bigl[%
S_{ab}+\lambda \frac{d-2}{4\left( d-1\right) }T_{ab}  \notag \\
&&-\frac{d-2}{8\left( d-1\right) }g_{ab}\lambda ^{2}\Bigr]\ .
\end{eqnarray}%
Here $S_{ab}$ denotes a quadratic expression in $T_{ab}$: 
\begin{eqnarray}
S_{ab} &=&\frac{1}{4}\Biggl[-T_{ac}^{\ }T_{b}^{c}+\frac{1}{d-1}TT_{ab} 
\notag \\
&&-\frac{g_{ab}}{2}\left( -T_{cd}^{\ }T^{cd}+\frac{1}{d-1}T^{2}\right) %
\Biggr]\ .  \label{S}
\end{eqnarray}%
By defining the brane gravitational constant and the brane cosmological
\textquotedblleft constant\textquotedblright\ through 
\begin{align}
\kappa ^{2}& =\frac{d-2}{4\left( d-1\right) }\widetilde{\kappa }^{4}\lambda
\ , \\
\Lambda & =\frac{\kappa ^{2}\lambda }{2}-\frac{\overline{L}}{d}-\widetilde{%
\kappa }^{2}\frac{d-2}{d}\overline{\left( n^{c}n^{d}\widetilde{\Pi }%
_{cd}\right) }\ ,  \label{Lambda}
\end{align}%
we obtain the effective Einstein equation (\ref{modEgen}). Among the source
terms we find $\overline{L}_{ab}$, which is defined as 
\begin{equation}
\overline{L}_{ab}=\overline{K}_{ab}\overline{K}-\overline{K}_{ac}\overline{K}%
_{b}^{c}-\frac{g_{ab}}{2}\left( \overline{K}^{2}-\overline{K}_{cd}\overline{K%
}^{cd}\right) \ ,
\end{equation}%
with $\overline{L}_{ab}^{TF}$ and $\overline{L}$ its tracefree part and
trace. Finally $\overline{\mathcal{P}}_{ab}$ is given by the pull-back of
the bulk energy-momentum tensor to the brane: 
\begin{equation}
\overline{\mathcal{P}}_{ab}=\widetilde{\kappa }^{2}\frac{d-2}{d-1}\overline{%
\left( g_{a}^{c}{}g_{b}^{d}{}\widetilde{\Pi }_{cd}\right) ^{TF}}\ .
\end{equation}%
The first four terms of the rhs of the effective Einstein equation are
well-known \cite{SMS}. They are the cosmological term, the ordinary brane
matter source term (dominant at low energies), a quadratic term in the brane
energy-momentum (relevant at high energies), and the bulk electric
Weyl-curvature contribution. The only modification up to here is the
possibility of a varying cosmological \textquotedblleft
constant\textquotedblright\ (it depends both on the projection $\overline{%
\left( n^{c}n^{d}\widetilde{\Pi }_{cd}\right) }$ of the bulk energy-momentum
tensor and on the embedding of the brane). In addition to these there are
two new terms. The first of them, $\overline{L}_{ab}^{TF}$ represents the
imprint of the particular way the time-evolving brane is bent into the bulk
from both sides. This contribution disappears in the $Z_{2}$-symmetric case
(as well as the contribution $\overline{L}$ to $\Lambda $). The last term, $%
\overline{\mathcal{P}}_{ab}$ arises from the projection of the bulk
energy-momentum tensor on the brane, and is traceless by definition.

In terms of $\lambda $ and $T_{ab}$, Eqs. (\ref{trm}), (\ref{vm}) and (\ref%
{trlessm})\ can be written as: 
\begin{subequations}
\label{syst1}
\begin{gather}
\Delta \left( n^{a}n^{b}\widetilde{\Pi }_{ab}\right) =-\lambda \overline{K}%
+T_{ab}\overline{K}^{ab}\ ,  \label{trm1} \\
\Delta \left( g_{a}^{c}{}n^{d}{}\widetilde{\Pi }_{cd}\right) =-\nabla
_{c}T_{a}^{c}\ ,  \label{vm1} \\
\widetilde{\kappa }^{2}\frac{d-2}{d-1}\Delta \left( g_{a}^{c}{}g_{b}^{d}{}%
\widetilde{\Pi }_{cd}\right) ^{TF}=\Delta \mathcal{E}_{ab}  \notag \\
+\widetilde{\kappa }^{2}\left[ \overline{K}T_{ab}+\frac{T}{d-1}\overline{K}%
_{ab}+\frac{d-2}{d-1}\lambda \overline{K}_{ab}-2\overline{K}_{(a}^{c}T_{b)c}%
\right] ^{TF}\ .  \label{trlessm1}
\end{gather}%
Thus Eqs. (\ref{vp}), (\ref{Edefpm}), (\ref{modEgen}) and (\ref{syst1}) are
the complete set of equations in the generic case of non-$Z_{2}$-symmetric
bulk and arbitrary energy-momentum tensors both on the brane and in the bulk.

The Bianchi identity in $d$ dimensions allows for the expression of the
longitudinal part of $\left( \overline{\mathcal{E}}_{ab}-\overline{L}%
_{ab}^{TF}-\overline{\mathcal{P}}_{ab}\right) $: 
\end{subequations}
\begin{gather}
\nabla ^{a}\left( \overline{\mathcal{E}}_{ab}-\overline{L}_{ab}^{TF}-%
\overline{\mathcal{P}}_{ab}\right) =  \notag \\
\frac{\nabla _{b}\overline{L}}{d}+\widetilde{\kappa }^{2}\frac{d-2}{d}\nabla
_{b}\overline{\left( n^{c}n^{d}\widetilde{\Pi }_{cd}\right) }-\kappa
^{2}\Delta \left( g_{b}^{c}{}n^{d}{}\widetilde{\Pi }_{cd}\right)  \notag \\
+\frac{\widetilde{\kappa }^{4}}{4}\left( T_{b}^{c}-\frac{T}{d-1}%
g_{b}^{c}\right) \Delta \left( g_{c}^{a}{}n^{d}{}\widetilde{\Pi }_{ad}\right)
\notag \\
+\frac{\widetilde{\kappa }^{4}}{4}\left[ 2T^{ac}\nabla _{\lbrack b}T_{a]c}+%
\frac{1}{d-1}\left( T_{ab}\nabla ^{a}T-T\nabla _{b}T\right) \right] \ .
\label{Bianchi}
\end{gather}%
This equation has important cosmological implications, as will be discussed
in the next section.

\section{Perfect fluid on Friedmann brane}

Friedmann branes, obeying cosmological symmetries are characterized by the
metric 
\begin{equation}
g_{ab}=-u_{a}u_{b}+a^{2}\left( \tau \right) h_{ab}\ ,  \label{fmetric}
\end{equation}%
where $a\left( \tau \right) $ is the scale factor and $h_{ab}$ is a $\left(
d-1\right) $ metric with \textit{constant} curvature (characterized by the
curvature index $k=1,0,-1$) of the maximally symmetric spacial slices with
constant $\tau $ . The timelike congruence $u^{a}=\left( \partial /\partial
\tau \right) ^{a}$ obeys $u^{a}u_{a}=-1$ and $h_{ab}u^{a}=0$. It is not
difficult to prove $u_{b}\nabla _{a}u^{b}=u^{b}\nabla _{b}u^{a}=0$. We
denote by a dot the time-derivative with respect to $\tau $, which in the
generic case is defined as the Lie-derivative in the $u^{a}$ direction,
projected into the hypersurface perpendicular to $u^{a}$ (of constant $\tau $%
). Then, from the condition $\dot{h}_{ab}=0$ we find 
\begin{equation}
u^{c}\nabla _{c}h_{ab}=-\frac{1}{a^{2}}\left( \nabla _{a}u_{b}+\nabla
_{b}u_{a}\right) \ ,
\end{equation}%
the trace of which implies $\nabla _{a}u^{a}=\left( d-1\right) \dot{a}/a$ .

When there is a perfect fluid on the brane, it has the energy-momentum
tensor 
\begin{equation}
T_{ab}=\rho \left( \tau \right) u_{a}u_{b}+p\left( \tau \right) a^{2}h_{ab}\
,  \label{fenmom}
\end{equation}%
with $u^{a}$ representing its $d$-velocity. Spatial isotropy and homogeneity
implies $h_{ab}\nabla ^{b}a=h_{ab}\nabla ^{b}\rho =h_{ab}\nabla ^{b}p=0$.

The quadratic term (\ref{S}) then becomes: 
\begin{equation}
\widetilde{\kappa }^{4}S_{ab}=\kappa ^{2}\frac{\rho }{\lambda }\left[ \frac{%
\rho }{2}u_{a}u_{b}+\left( \frac{\rho }{2}+p\right) a^{2}h_{ab}\right] \ .
\label{fS}
\end{equation}%
We complete the bookkeeping of the source terms by introducing the effective
non-local energy density $U$ arising from the totality of non-local \textit{%
tracefree} terms in the effective Einstein equation (\ref{modEgen}): 
\begin{equation}
-\overline{\mathcal{E}}_{ab}+\overline{L}_{ab}^{TF}+\overline{\mathcal{P}}%
_{ab}=\kappa ^{2}U\left( u_{a}u_{b}+\frac{a^{2}}{d-1}h_{ab}\right) \ .
\label{U}
\end{equation}%
$U$ is a generalization of the effective non-local energy density introduced
in the case of $Z_{2}$ symmetric, cosmological bulk \cite{Maartens2}.

Next we specify the system (\ref{syst1}) for the perfect fluid
energy-momentum tensor. Eqs. (\ref{trm1}) and (\ref{trlessm1}) give 
\begin{gather}
\Delta \left( n^{a}n^{b}\widetilde{\Pi }_{ab}\right) =\left( p-\lambda
\right) \overline{K}+\left( \rho +p\right) u_{a}u_{b}\overline{K}^{ab}\ ,
\label{trm2} \\
\widetilde{\kappa }^{2}\frac{d-2}{d-1}\Delta \left( g_{a}^{c}{}g_{b}^{d}{}%
\widetilde{\Pi }_{cd}\right) ^{TF}=\Delta \mathcal{E}_{ab}  \notag \\
+\widetilde{\kappa }^{2}\Biggl[\left( \rho +p\right) \overline{K}%
u_{a}u_{b}-2\left( \rho +p\right) u_{(a}\overline{K}_{b)}^{c}u_{c}  \notag \\
-\frac{\rho +\left( d-1\right) p-\left( d-2\right) \lambda }{d-1}\overline{K}%
_{ab}\Biggr]^{TF}\ ,  \label{trlessm2}
\end{gather}%
while Eq. (\ref{vm1}) decouples into the following time- and space
components: 
\begin{gather}
\Delta \left( u^{c}{}n^{d}{}\widetilde{\Pi }_{cd}\right) =\dot{\rho}+\left(
d-1\right) \frac{\dot{a}}{a}\left( \rho +p\right) \ ,  \label{rhodot} \\
h_{ab}\Delta \left( g^{ac}{}n^{d}{}\widetilde{\Pi }_{cd}\right) =0\ .
\label{piproj}
\end{gather}%
By virtue of Eqs. (\ref{U}), (\ref{rhodot}) and (\ref{piproj}) the space
projection of the Bianchi identity (\ref{Bianchi}) trivially vanishes. (We
also use in the proof that in order to fulfill the cosmological symmetries, $%
U$ as well as $\overline{L}$ and $\overline{\left( n^{c}n^{d}\widetilde{\Pi }%
_{cd}\right) }$, (both contributing to $\Lambda $) are pure time-functions,
i.e. they have vanishing spatial derivatives.) The time projection of the
Bianchi identity (\ref{Bianchi}) gives: 
\begin{eqnarray}
\kappa ^{2}\left( \dot{U}+d\frac{\dot{a}}{a}U\right) &=&\frac{1}{d}\left[ 
\widetilde{\kappa }^{2}\left( d-2\right) \overline{\left( n^{c}n^{d}%
\widetilde{\Pi }_{cd}\right) }+\overline{L}\right] ^{\mathbf{\skew{01}
{\dot} {}}}  \notag \\
&&-\kappa ^{2}\left( 1+\frac{\rho }{\lambda }\right) \Delta \left(
u^{c}{}n^{d}{}\widetilde{\Pi }_{cd}\right) \ .  \label{Bianchi1}
\end{eqnarray}%
The homogeneous part of the above equation integrates to 
\begin{equation}
U=U_{0}\left( \frac{a_{0}}{a}\right) ^{d}\ ,  \label{Usol1}
\end{equation}%
where $U_{0}a_{0}^{d}$ is an integration constant. Variation of the constant 
$U_{0}$ gives a first order ordinary differential equation: 
\begin{equation}
\kappa ^{2}\left( \frac{a_{0}}{a}\right) ^{d}\dot{U}_{0}+\dot{\Lambda}%
+\kappa ^{2}\left( 1+\frac{\rho }{\lambda }\right) \Delta \left(
u^{c}{}n^{d}{}\widetilde{\Pi }_{cd}\right) =0\ .  \label{U0}
\end{equation}%
(We have employed the definition of $\Lambda $ given in Eq. (\ref{Lambda})).
The solution of Eq. (\ref{U0}) depends both on the bulk matter and on the
details of the embedding of the brane in the bulk.

Finally we compute the Einstein tensor: 
\begin{eqnarray}
G_{ab} &=&\frac{\left( d-2\right) }{2}\Biggl\{\left( d-1\right) \frac{\dot{a}%
^{2}+k}{a^{2}}u_{a}u_{b}  \notag \\
&&-\left[ 2a\ddot{a}+\left( d-3\right) \left( \dot{a}^{2}+k\right) \right]
h_{ab}\Biggr\}\ ,  \label{fein}
\end{eqnarray}%
which is the last piece of information required in order to write the
effective Einstein equation (\ref{modEgen}.) Its non-trivial projections
combine to give the generalized Friedmann and generalized Raychaudhuri
equations: 
\begin{gather}
\left( d-1\right) \left( d-2\right) \frac{\dot{a}^{2}+k}{a^{2}}=2\Lambda
+2\kappa ^{2}\rho \left( 1+\frac{\rho }{2\lambda }\right)  \notag \\
+2\kappa ^{2}U_{0}\left( \frac{a_{0}}{a}\right) ^{d}\ ,  \label{Fried} \\
\left( d-1\right) \left( d-2\right) \frac{\ddot{a}}{a}=2\Lambda -\kappa ^{2}%
\Biggl\{\left[ d-3+\left( d-2\right) \frac{\rho }{\lambda }\right] \rho 
\notag \\
+\left( d-1\right) \left( 1+\frac{\rho }{\lambda }\right) p\Biggr\}-\left(
d-2\right) \kappa ^{2}U_{0}\left( \frac{a_{0}}{a}\right) ^{d}\ .
\label{Raych}
\end{gather}%
Apart from the dimension-carrying index $d,$ at first glance the above
equations are\textit{\ identical} with the corresponding equations of \cite%
{BDEL} and \cite{Maartens}, obtained for $Z_{2}$ symmetric cosmological
bulk. Still, important differences arise from the non-constant character of $%
\Lambda $ and $U_{0},$ given in the generic case by Eqs. (\ref{Lambda}) and (%
\ref{U0}). Another distinctive feature is that the $U_{0}$ term cannot be
interpreted any more as pure dark radiation. A glance at Eqs. (\ref{U}) and (%
\ref{Usol1}) shows that it carries both radiative degrees of freedom (from
the electric bulk Weyl tensor), as well as imprints of the bulk matter and
of the particular way the brane is bent into the left and right bulk
regions. The equation (\ref{U0}) defining the potential $U_{0}$ is an
integrability condition, which can be equivalently derived by taking the
time derivative of the generalized Friedmann equation, then employing Eq. (%
\ref{rhodot}) to eliminate $\dot{\rho}$ and Eqs. (\ref{Fried}) and (\ref%
{Raych}) to eliminate all derivatives of $a$.

Let us recall, that the information on the gravitational field is completed
by Eqs. (\ref{vp}) and (\ref{trm2}) (these represent $\left( d+1\right) $
constraints on the mean extrinsic curvature and bulk matter), Eqs. (\ref%
{rhodot}) and (\ref{piproj}) ($d$ constraints on the bulk matter), Eq. (\ref%
{trlessm2}) determining $\Delta \mathcal{E}_{ab}$ and Eq. (\ref{Edefpm})
giving $E^{\pm }$. Also, the Lanczos equation 
\begin{equation}
\Delta K_{ab}\!=\!-\widetilde{\kappa }^{2}\!\left[ \!\left( \frac{d-2}{d-1}%
\rho +p-\frac{\lambda }{d-1}\right) \!u_{a}u_{b}+\frac{\rho +\lambda }{d-1}%
a^{2}h_{ab}\right]  \label{Lancz}
\end{equation}%
remains a useful link between dynamical and geometrical quantities defined
on the brane.

We summarize the above results in the Appendix for $d=4$. The algorithmic
way the equations are grouped is meant to facilitate the search for non-$%
Z_{2}$-symmetric FRW brane-world solutions. The equations governing
off-brane evolution are also presented there.

In the following two sections we apply these generic equations for the 
\textit{five-dimensional} Reissner-Nordstr\"{o}m-Anti de Sitter bulk and
charged Vaidya-Anti de Sitter bulk, both containing a \textit{%
four-dimensional} Friedmann brane. On the two sides of the brane the bulk is
characterized by different mass and charge functions, also by different
cosmological constants. As the generalized Friedmann and Raychaudhuri
equations are given in a form invariant with respect to different choices of
the bulk and the embeddings, only the complementary set of equations (\ref%
{vp}), (\ref{Edefpm}), and (\ref{trm2})-(\ref{piproj}) will change from case
to case.

\section{Reissner-Nordstr\"{o}m-Anti de Sitter bulk}

The generalization of the 4-dimensional \ Reissner-Nordstr\"{o}m solution to
a cosmological context in 5 dimensions was discussed in \cite{BVisser}. The
metric 
\begin{align}
d\widetilde{s}^{2}& =-f\left( r;k\right) dt^{2}+\frac{dr^{2}}{f\left(
r;k\right) }  \notag \\
& +r^{2}\left[ d\chi ^{2}+\mathcal{H}^{2}\left( \chi ;k\right) \left(
d\theta ^{2}+\sin ^{2}\theta d\phi ^{2}\right) \right] \ ,  \label{RNAdS5}
\end{align}%
with 
\begin{equation}
\mathcal{H}\left( \chi ;k\right) =\left\{ 
\begin{array}{c}
\sin \chi \ ,\qquad k=1 \\ 
\chi \ ,\qquad k=0 \\ 
\sinh \chi \ ,\qquad k=-1%
\end{array}%
\right. \ ,
\end{equation}%
can be also written as 
\begin{equation}
\widetilde{g}_{ab}=-u_{a}u_{b}+n_{a}n_{b}+r^{2}h_{ab}\ ,  \label{RNAdS5a}
\end{equation}%
where a time coordinate $\tau $ was introduced through the timelike vector 
\begin{equation}
u=\frac{\partial }{\partial \tau }=\dot{t}\frac{\partial }{\partial t}+\dot{r%
}\frac{\partial }{\partial r}\ ,
\end{equation}%
with unit negative norm, implying 
\begin{equation}
f\dot{t}=\left( \dot{r}^{2}+f\right) ^{1/2}\ .  \label{tdotRN}
\end{equation}%
We have chosen the $\dot{t}>0$ root and a dot denotes derivatives with
respect to $\tau $. Then the unit normal 1-form to both $h_{ab}$ and $u_{a}$
is determined up to a sign: 
\begin{equation}
n=\pm \left( -1\right) ^{\sigma }\left( -\dot{r}dt+\dot{t}dr\right) \ .
\label{normalRN}
\end{equation}%
The $+$ sign refers to right-pointing normal. The final results will not
depend on this choice of the orientation. We have inserted an additional
sign $\left( -1\right) ^{\sigma }$ to allow the \textit{outgoing} coordinate 
$y$, defined by%
\begin{equation}
n=\pm \left( -1\right) ^{\sigma }dy\ ,
\end{equation}%
to increase either in the right or left directions. We will specify later
the meaning of the exponent $\sigma $. Finally the 1-form field $u_{a}$ is%
\begin{equation}
u=-\left( \dot{r}^{2}+f\right) ^{1/2}dt+\dot{a}dr=-d\tau \ .
\label{tangentRN}
\end{equation}

If the bulk contains an electromagnetic field characterized by the potential
1-form%
\begin{equation}
A=\frac{q}{r^{2}}dt\ ,
\end{equation}%
its energy-momentum tensor will be 
\begin{equation}
\widetilde{T}_{ab}^{EM}=\frac{3q^{2}}{r^{6}}\left(
u_{a}u_{b}-n_{a}n_{b}+r^{2}h_{ab}\right) \ .  \label{TEMRN}
\end{equation}%
Then the bulk Einstein equation for the metric ansatz (\ref{RNAdS5}) with
the source term%
\begin{equation}
\widetilde{\Pi }_{ab}=-\widetilde{\Lambda }\widetilde{g}_{ab}+\widetilde{T}%
_{ab}^{EM}\ 
\end{equation}%
is satisfied for%
\begin{equation}
f\left( r;k\right) =k-\frac{2m}{r^{2}}-\frac{\widetilde{\kappa }^{2}%
\widetilde{\Lambda }}{6}r^{2}+\frac{\widetilde{\kappa }^{2}q^{2}}{r^{4}}\ ,
\label{ffRN}
\end{equation}%
$m$ and $q$ being the mass and charge of the central black hole (or
\textquotedblleft stellar object\textquotedblright , in the absence of a
horizon) and $\widetilde{\Lambda }<0$ the bulk cosmological constant. In
principle, all of these constants can take different values on the two sides
of the brane. Thus we will drop the assumption of $Z_{2}$-symmetry and
obtain more generic results than in \cite{BVisser}.

If one passes from the coordinates $\left( t,r\right) $ to $\left( \tau
,y\right) $, the position of the brane can be simply specified as $y=$const.
This choice is equivalent to the embedding relations $t=t\left( \tau \right) 
$ and $r=a\left( \tau \right) $. Therefore we replace $\left( r,\ \dot{r}%
\right) $\ with $\left( a,\ \dot{a}\right) $. As for the embedding relation $%
t=t\left( \tau \right) $, we know only Eq. (\ref{tdotRN}). By construction, $%
n$ is the normal and $u$ the tangent vector to the brane. Thus we take for
the induced metric the expression (\ref{fmetric}). By this we have assumed
that the bulk has the spatial symmetries of the brane \footnote{%
For such a brane and pure cosmological bulk a \textquotedblright generalized
Birkhoff theorem\textquotedblright\ holds \cite{BCG}, stating that the bulk
is the 5 dimensional Schwarzschild-anti de Sitter space-time, a particular
case in our treatment. However when the brane has the additional static
symmetry (Einstein brane), the derivation of the above mentioned\
\textquotedblleft generalized Birkhoff theorem\textquotedblright\ is
obstructed, and other bulk solutions are possible, as shown in \cite%
{Einbrane}.}.

The extrinsic curvature for such hypersurfaces is:%
\begin{eqnarray}
K_{ab} &=&\mp \left( -1\right) ^{\sigma }\Biggl[\frac{\ddot{a}+\frac{1}{2}%
\frac{\partial f}{\partial a}}{\left( \dot{a}^{2}+f\right) ^{1/2}}u_{a}u_{b}
\notag \\
&&-\left( \dot{a}^{2}+f\right) ^{1/2}ah_{ab}\Biggr]\ .
\end{eqnarray}%
$K_{ab}$ on the two sides depends on the actual value of the function $f$
and the sign ambiguity arises from the ambiguity in the choice of the normal.

As we have not imposed the $Z_{2}$-symmetry, at this point we have to raise
the question, whether inner or outer regions of the Reissner-Nordstr\"{o}%
m-Anti de Sitter space-time will be glued together. We introduce a pair of
indices $\eta _{R,L}$ which take the value $1$ for inner regions and $0$ for
outer regions. Then, according to our definitions%
\begin{equation}
\sigma =\left\{ 
\begin{array}{c}
\eta _{R}\ ,\qquad \text{right region} \\ 
\eta _{L}+1\ ,\qquad \text{left region}%
\end{array}%
\right. \ .
\end{equation}%
Then the extrinsic curvatures on the two sides of the brane are%
\begin{eqnarray}
K_{ab}^{R} &=&\mp \left( \frac{A_{R}}{B_{R}}u_{a}u_{b}-B_{R}ah_{ab}\right) \
,  \notag \\
K_{ab}^{L} &=&\pm \left( \frac{A_{L}}{B_{L}}u_{a}u_{b}-B_{L}ah_{ab}\right) \
,
\end{eqnarray}%
where $R,L$ refer to right and left regions, respectively, and the following
notations were introduced for convenience $\left( I=R,L\right) $:%
\begin{eqnarray}
A_{I} &=&\ddot{a}+\frac{1}{2}\frac{\partial f_{I}}{\partial a}\ ,
\label{ARN} \\
B_{I} &=&\left( -1\right) ^{\eta _{I}}\left( \dot{a}^{2}+f_{I}\right)
^{1/2}\ .  \label{B}
\end{eqnarray}%
Then, according to the definition of the jump and mean value of the
extrinsic curvature: 
\begin{eqnarray}
\Delta K_{ab}\!\!\! &=&\!\!\!-\!\left[ \!\left( \frac{A_{R}}{B_{R}}+\frac{%
A_{L}}{B_{L}}\right) u_{a}u_{b}-\!\left( B_{R}+B_{L}\right) ah_{ab}\right] \
,  \notag \\
2\overline{K}_{ab}\!\!\! &=&\!\!\!\mp \!\left[ \!\left( \frac{A_{R}}{B_{R}}-%
\frac{A_{L}}{B_{L}}\right) u_{a}u_{b}-\!\left( B_{R}-B_{L}\right) ah_{ab}%
\right] .
\end{eqnarray}%
We also have 
\begin{equation}
\frac{2\overline{L}}{3}=-\frac{B_{R}-B_{L}}{a}\left[ \frac{A_{R}}{B_{R}}-%
\frac{A_{L}}{B_{L}}+\frac{B_{R}-B_{L}}{a}\right] \ .
\end{equation}%
The equations (\ref{A1}) and (\ref{A3}) are then satisfied, while Eq. (\ref%
{A2}) gives%
\begin{equation}
\dot{\rho}+3\frac{\dot{a}}{a}\left( \rho +p\right) =0\ .
\end{equation}%
This first order differential relation among the scale factor, pressure and
density of the perfect fluid (being independent both of $\Lambda $ and $%
U_{0} $) guarantees that for a given equation of state $p\left( \rho \right) 
$, the density has the standard expression as function of the scale factor.

We actually do not have to carry out in full detail the program described in
the Appendix. This is because our choices of the bulk and brane metrics
already constrain the embedding, leading to the above extrinsic curvature.
Then a shortcut would be to employ Eqs. (\ref{A4}) and (\ref{A12}) in order
to express $\dot{a}^{2}$ and $\ddot{a}$ algebraically. (As the sign
ambiguity was verified to cancel out from all equations, from now on we mean
by $\Delta $ the differences between quantities taken from the $R$ and $\ L$
regions.) First we find by pure algebra that 
\begin{eqnarray}
&&\overline{B}=-\frac{\widetilde{\kappa }^{2}a}{6}\left( \rho +\lambda
\right) \ ,  \notag \\
&&\Delta B=\frac{3\Delta A+\widetilde{\kappa }^{2}aC}{\widetilde{\kappa }%
^{2}\left( \rho +\lambda \right) }\ ,
\end{eqnarray}%
then 
\begin{eqnarray}
\frac{\dot{a}^{2}+\overline{f}}{a^{2}} &=&\frac{\kappa ^{2}\lambda }{6}+%
\frac{\kappa ^{2}\rho }{3}\left( 1+\frac{\rho }{2\lambda }\right) +\frac{%
\left( \Delta B\right) ^{2}}{4a^{2}}\ , \\
\frac{\overline{A}}{a} &=&\frac{\kappa ^{2}\lambda }{6}-\frac{\kappa ^{2}}{6}%
\left[ \rho \left( 1+\frac{2\rho }{\lambda }\right) +3p\left( 1+\frac{\rho }{%
\lambda }\right) \right]  \notag \\
&&+\frac{C\Delta B}{2a\left( \rho +\lambda \right) }+\frac{3\left( p-\lambda
\right) \left( \Delta B\right) ^{2}}{4a^{2}\left( \rho +\lambda \right) }\ .
\end{eqnarray}%
By $C$ we have denoted the source term in Eq. (\ref{A4}), multiplied by $\mp 
$, in the present case 
\begin{equation}
C=\frac{6\overline{q}\Delta q}{a^{6}}+\Delta \widetilde{\Lambda }\ .
\label{CRN}
\end{equation}%
By employing the definition of the metric function (\ref{ffRN}), and Eqs. (%
\ref{ARN}), (\ref{CRN}), we obtain 
\begin{equation}
\Delta B=\frac{12a^{2}\Delta m-12\widetilde{\kappa }^{2}\overline{q}\Delta q+%
\widetilde{\kappa }^{2}a^{6}\Delta \widetilde{\Lambda }}{2\widetilde{\kappa }%
^{2}a^{5}\left( \rho +\lambda \right) }\ ,  \label{delB}
\end{equation}%
and the generalized Friedmann and Raychaudhuri equations%
\begin{gather}
\frac{\dot{a}^{2}+k}{a^{2}}=\frac{\Lambda _{0}}{3}+\frac{\kappa ^{2}\rho }{3}%
\left( 1+\frac{\rho }{2\lambda }\right) +\frac{2\overline{m}}{a^{4}}-\frac{%
\widetilde{\kappa }^{2}\overline{q}^{2}}{a^{6}}  \notag \\
-\frac{\widetilde{\kappa }^{2}\left( \Delta q\right) ^{2}}{4a^{6}}+\frac{%
\left( \Delta B\right) ^{2}}{4a^{2}}\ ,  \label{Fr} \\
\frac{\ddot{a}}{a}=\frac{\Lambda _{0}}{3}-\frac{\kappa ^{2}}{6}\left[ \rho
\left( 1+\frac{2\rho }{\lambda }\right) +3p\left( 1+\frac{\rho }{\lambda }%
\right) \right]  \notag \\
-\frac{2\overline{m}}{a^{4}}+\frac{2\widetilde{\kappa }^{2}\overline{q}^{2}}{%
a^{6}}  \notag \\
+\frac{\widetilde{\kappa }^{2}\left( \Delta q\right) ^{2}}{2a^{6}}+\frac{%
C\Delta B}{2a\left( \rho +\lambda \right) }+\frac{3\left( p-\lambda \right)
\left( \Delta B\right) ^{2}}{4a^{2}\left( \rho +\lambda \right) }\ ,
\label{RRN}
\end{gather}%
where $\Lambda _{0}$ is a true constant given as: 
\begin{equation}
2\Lambda _{0}=\kappa ^{2}\lambda +\widetilde{\kappa }^{2}\overline{%
\widetilde{\Lambda }}\ .
\end{equation}%
A comparison with Eqs. (\ref{A9}) and (\ref{A10}) gives the cosmological
\textquotedblleft constant\textquotedblright\ and the potential $U_{0}$:%
\begin{gather}
\frac{\Lambda }{3}=\frac{\Lambda _{0}}{3}+\frac{\widetilde{\kappa }^{2}%
\overline{q}^{2}}{2a^{6}}+\frac{\widetilde{\kappa }^{2}\left( \Delta
q\right) ^{2}}{8a^{6}}  \notag \\
+\frac{C\Delta B}{4a\left( \rho +\lambda \right) }+\frac{\left( \rho
+3p-2\lambda \right) \left( \Delta B\right) ^{2}}{8a^{2}\left( \rho +\lambda
\right) }\ , \\
\frac{\kappa ^{2}}{3}U_{0}\left( \frac{a_{0}}{a}\right) ^{4}=\frac{2%
\overline{m}}{a^{4}}-\frac{3\widetilde{\kappa }^{2}\overline{q}^{2}}{2a^{6}}-%
\frac{3\widetilde{\kappa }^{2}\left( \Delta q\right) ^{2}}{8a^{6}}  \notag \\
-\frac{C\Delta B}{4a\left( \rho +\lambda \right) }+\frac{\left( \rho
-3p+4\lambda \right) \left( \Delta B\right) ^{2}}{8a^{2}\left( \rho +\lambda
\right) }\ .
\end{gather}%
The Friedmann equation (\ref{Fr}) and Raychaudhuri equation (\ref{RRN}),
after suitable conversion of notation, reduce to the corresponding results
of \cite{Stoica}, \cite{BCG} in the case $q=0$. When the cosmological
constant is the same in both bulk regions ($\Delta \widetilde{\Lambda }=0$),
earlier results \cite{Ida} are recovered.

In the $Z_{2}$-symmetric limit we recover the Friedmann equation given in 
\cite{BVisser}. In the absence of charge, the Friedmann and Raychaudhuri
equations given in \cite{BDEL} and \cite{Maartens} emerge.

\section{Charged Vaidya-Anti de Sitter bulk}

The generalization of the 4-dimensional \ charged Vaidya solution \cite%
{VaidyaBonnor} in a cosmological context was discussed in \cite{WW}. We will
do the same here in 5 dimensions. Let us start with the bulk metric written
in Eddington-Finkelstein type coordinates,%
\begin{align}
d\widetilde{s}^{2}& =-f\left( v,r;k\right) dv^{2}+2\epsilon dvdr  \notag \\
& +r^{2}\left[ d\chi ^{2}+\mathcal{H}^{2}\left( \chi ;k\right) \left(
d\theta ^{2}+\sin ^{2}\theta d\phi ^{2}\right) \right] \ ,  \label{ChVAdS5}
\end{align}%
where $\epsilon =1$ holds for an outgoing $v$ coordinate (the $v=$const.
lines are ingoing), while $\epsilon =-1$ for ingoing $v$ ($v=$const. lines
outgoing). It can also be written as 
\begin{equation}
\widetilde{g}_{ab}=-u_{a}u_{b}+n_{a}n_{b}+r^{2}h_{ab}\ ,  \label{ChVAdS5a}
\end{equation}%
where%
\begin{equation}
u=\frac{\partial }{\partial \tau }=\dot{v}\frac{\partial }{\partial v}+\dot{r%
}\frac{\partial }{\partial r}\ ,
\end{equation}%
has unit negative norm, implying 
\begin{equation}
f\dot{v}=\epsilon \dot{r}+\left( \dot{r}^{2}+f\right) ^{1/2}\ .  \label{vdot}
\end{equation}%
We have again chosen the $\dot{v}>0$ root and a dot again denotes
derivatives with respect to $\tau $. Then the unit normal 1-form to both $%
h_{ab}$ and $u_{a}$ becomes: 
\begin{equation}
n=\pm \left( -1\right) ^{\sigma }\left( -\dot{r}dv+\dot{v}dr\right) \ .
\label{normal}
\end{equation}%
Finally the 1-form field $u_{a}$ is 
\begin{equation}
u=-\left( \dot{r}^{2}+f\right) ^{1/2}dv+\epsilon \dot{v}dr=-d\tau \ .
\label{tangent}
\end{equation}

We suppose that the bulk contains radiation (geometrical optics limit: null
dust) with energy-momentum tensor%
\begin{equation}
\widetilde{T}_{ab}^{ND}=\frac{3\beta \left( v,r\right) }{\widetilde{\kappa }%
^{2}r^{3}}l_{a}l_{b}\ .  \label{TND}
\end{equation}%
Here $\beta \left( v,r\right) $ determines the energy density (it has the
dimension of a linear density of mass) and $l$ is a null 1-form: 
\begin{equation}
l=dv=\dot{v}\left[ \pm \epsilon \left( -1\right) ^{\sigma }n-u\right] \ .
\end{equation}%
Such radiation is ingoing for $\epsilon =1$ and outgoing for $\epsilon =-1$.
In the bulk there is also an electromagnetic contribution, 
\begin{equation}
\widetilde{T}_{ab}^{EM}=\frac{3q^{2}\left( v\right) }{r^{6}}\left(
u_{a}u_{b}-n_{a}n_{b}+r^{2}h_{ab}\right) \ ,  \label{TEM}
\end{equation}%
generated by a null 5-potential 
\begin{equation}
A_{a}=\frac{q\left( v\right) }{r^{2}}l_{a}\ .
\end{equation}

Then the bulk Einstein equation for the metric ansatz (\ref{ChVAdS5}), with
the source term%
\begin{equation}
\widetilde{\Pi }_{cd}=-\widetilde{\Lambda }\widetilde{g}_{ab}+\widetilde{T}%
_{ab}^{ND}+\widetilde{T}_{ab}^{EM}\ ,
\end{equation}%
is solved by%
\begin{equation}
\epsilon \beta =\frac{dm}{dv}-\frac{\widetilde{\kappa }^{2}q}{r^{2}}\frac{dq%
}{dv}\ ,  \label{beta}
\end{equation}%
and%
\begin{equation}
f\left( v,r;k\right) =k-\frac{1}{r^{2}}\left[ 2m\left( v\right) +\frac{%
\widetilde{\kappa }^{2}\widetilde{\Lambda }}{6}r^{4}-\frac{\widetilde{\kappa 
}^{2}q^{2}\left( v\right) }{r^{2}}\right] \ .  \label{ff}
\end{equation}%
The functions $m\left( v\right) $ and $q\left( v\right) $ are freely
specifiable.

The brane is given by the embedding relations $v=v\left( \tau \right) $
(through Eq. (\ref{vdot})) and $r=a\left( \tau \right) $, thus we replace $%
\left( r,\ \dot{r}\right) $\ with $\left( a,\ \dot{a}\right) $ in the above
formulae. By construction, $n$ is the normal to the brane. The induced
metric is (\ref{fmetric}) and the extrinsic curvature becomes 
\begin{eqnarray}
K_{ab} &=&\mp \left( -1\right) ^{\sigma }\Biggl[\frac{2\ddot{a}+\frac{%
\partial f}{\partial a}-\epsilon \dot{v}^{2}\frac{\partial f}{\partial v}}{%
2\left( \dot{a}^{2}+f\right) ^{1/2}}u_{a}u_{b}  \notag \\
&&-\left( \dot{a}^{2}+f\right) ^{1/2}ah_{ab}\Biggr]\ .
\end{eqnarray}%
We note that the differences with respect to the Reissner-Nordstr\"{o}m-Anti
de Sitter case arise in the definition of the function $A$: 
\begin{equation}
2A_{I}=2\ddot{a}+\frac{\partial f_{I}}{\partial a}-\epsilon _{I}\dot{v}%
_{I}^{2}\frac{\partial f_{I}}{\partial v}\ ,  \label{AV}
\end{equation}%
and in an additional term contained by $C$:%
\begin{equation}
C=\frac{6\overline{q}\Delta q}{a^{6}}+\Delta \widetilde{\Lambda }-\frac{%
3\Delta \left( \beta \dot{v}^{2}\right) }{\widetilde{\kappa }^{2}a^{3}}\ .
\label{CV}
\end{equation}%
Since, by virtue of Eq. (\ref{beta}) the condition $2\epsilon _{I}\beta
_{I}+a^{2}\partial f_{I}/\partial v=0$ holds, the expression (\ref{delB})
for $\Delta B$, as well as the generalized Friedmann equation (\ref{Fr}),
are unchanged relative to the Reissner-Nordstr\"{o}m-Anti de Sitter case.
The Raychaudhuri equation acquires two new terms on the right hand side.
These are%
\begin{equation}
2\mathcal{K}=-\frac{\overline{\beta \dot{v}^{2}}}{a^{3}}-\frac{3\Delta B}{2%
\widetilde{\kappa }^{2}a^{4}\left( \rho +\lambda \right) }\Delta \left(
\beta \dot{v}^{2}\right) \ .  \label{RVcorr}
\end{equation}%
Then the cosmological \textquotedblleft constant\textquotedblright\ and the
potential in the charged Vaidya-Anti de Sitter case are found immediately
from the ones characterizing the Reissner-Nordstr\"{o}m-Anti de Sitter case: 
\begin{eqnarray}
\left( \frac{\Lambda }{3}\right) _{chVAdS5}\! &=&\!\left( \frac{\Lambda }{3}%
\right) _{RNAdS5}+\mathcal{K}\ , \\
\left( U_{0}\right) _{chVAdS5}\! &=&\!\left( U_{0}\right) _{RNAdS5}\!-\!%
\frac{3}{\kappa ^{2}}\!\!\!\!\left( \frac{a}{a_{0}}\right) ^{4}\!\mathcal{K}%
\ .
\end{eqnarray}%
(Remember that $\Lambda $ and $U_{0}$ on the r.h.s. should be computed with $%
C$ given by Eq. (\ref{CV}).) In the $Z_{2}$-symmetric limit the Friedmann
equation reduces to the one given in \cite{ChKN}. \qquad

The equations (\ref{A1}) and (\ref{A3}) are again satisfied, while Eq. (\ref%
{A2}) this time differs from the ordinary continuity equation:%
\begin{eqnarray}
\dot{\rho}+3\frac{\dot{a}}{a}\left( \rho +p\right) &=&\frac{3}{\widetilde{%
\kappa }^{2}a^{3}}\Delta \left[ \epsilon \left( -1\right) ^{\sigma }\beta 
\dot{v}^{2}\right]  \notag \\
&=&\frac{3}{\widetilde{\kappa }^{2}a^{3}}\sum\limits_{I=L,R}\epsilon
_{I}\left( -1\right) ^{\eta _{I}}\beta _{I}\dot{v}_{I}^{2}\ ,
\label{noconti}
\end{eqnarray}%
The global sign $\epsilon _{I}\left( -1\right) ^{\eta _{I}}$ of the two
terms in the sum above is negative for radiation leaving the brane and
positive for radiation arriving to the brane. For a given equation of state $%
p\left( \rho \right) $, this time the expression $\rho =\rho \left( a\right) 
$ is different from the standard one. This is due to the fact that the brane
radiates or is irradiated; thus there is no brane-energy conservation. (The $%
Z_{2}$-symmetric, uncharged limit of the case $\eta _{L}=\eta
_{R}=1=\epsilon _{L}=\epsilon _{R}$ was discussed in \cite{LSR}.)

Obviously, the expression $\beta \dot{v}^{2}$ is needed for the last four
equations. Of course, $\beta $ depends on the freely specifiable functions $%
m\left( v\right) $ and $q\left( v\right) $, as given by Eq. (\ref{beta}),
while $\dot{v}$ is determined by Eqs. (\ref{B}) and (\ref{vdot}). At this
point it is useful to note, that there is no natural normalization condition
for a null vector, therefore $l$ can be freely rescaled $l\rightarrow \sigma
l$ at the price that $\beta \rightarrow \beta /\sigma ^{2}$ is also rescaled
accordingly. Employing this freedom one can even shift all information about
the null dust into the null vector, by choosing $\sigma =\sqrt{\beta }$.
Alternatively, in the present case a simple way to proceed would be to
choose $\sigma =\dot{v}^{-1}$ which has the consequence that the new linear
density of mass is 
\begin{equation}
\alpha =\beta \dot{v}^{2}\ .  \label{alpha}
\end{equation}
Then one can interpret $\alpha $ in Eqs. (\ref{RVcorr})-(\ref{noconti}) as a
freely specifiable parameter. This choice was followed in \cite{LSR}

However we will follow a third route, guided by the desire to have freely
imposable functions with obvious meaning on the brane. The arbitrary metric
functions $m\left( v\right) $ and $q\left( v\right) $, due to the embedding
relation $v=v\left( \tau \right) $, can already be interpreted from a brane
point of view as arbitrary functions of time $m\left( \tau \right) $ and $%
q\left( \tau \right) $. Their time-derivatives, defined as $\dot{m}=\dot{v}%
dm/dv$ and $\dot{q}=\dot{v}dq/dv$, have again a natural interpretation for
an observer living on the brane. One can therefore define a third linear
density of mass by choosing $\sigma =\left( \dot{v}\right) ^{-1/2}$ as 
\begin{equation}
\gamma =\beta \dot{v}=\epsilon \left( \dot{m}-\frac{\widetilde{\kappa }^{2}q%
}{a^{2}}\dot{q}\right) \ ,
\end{equation}%
and rewrite Eqs. (\ref{RVcorr})-(\ref{noconti}) in terms of $\gamma $, which
has immediate interpretation for a brane observer. Then $\dot{v}$ arises
linearly in the term $\beta \dot{v}^{2}=\gamma \dot{v}$ and we obtain the
following expression:%
\begin{equation}
\beta _{I}\dot{v}_{I}^{2}=\dot{a}\frac{\gamma _{I}}{f_{I}}+\epsilon
_{I}\left( -1\right) ^{\eta _{I}}\frac{\gamma _{I}B_{I}}{f_{I}}\ .
\end{equation}%
This can be expressed in terms of $\overline{f}$, $\overline{B}$, $\overline{%
\gamma }$, $\Delta f$, $\Delta B$ and $\Delta \gamma $ or equivalently in
terms of average values and jumps of $m$, $\dot{m}$, $q$, $\dot{q}$ and $%
\widetilde{\Lambda }$. If this latter interpretation is chosen, an
interesting feature which emerges, is the occurrence of an $\dot{a}$ term in
both the Raychaudhuri and continuity equations. This can be avoided in the
first choice (\ref{alpha}), but then the relation between $\alpha $ and the
set $dm/dv$, $dq/dv$ will contain it.

\section{Concluding remarks}

We have given a generic decomposition of the Einstein equations, in which
the tensorial, vectorial and scalar projections are equivalent to the
effective Einstein, the Codazzi and the twice contracted Gauss equation. The
junction conditions applied across a brane separating two non-identical
spacetimes give rise to the final form of these equations. The effective
Einstein equation contains a varying cosmological constant, and extra terms
beyond the standard $Z_{2}$-symmetric case, which characterize the
non-symmetric embedding and the bulk matter.

The formalism can be applied for any situation. Of particular interest would
be the cases of branes containing black holes or obeying cosmological
symmetries. We have discussed the latter case here.

When the brane has cosmological symmetries and a perfect fluid, obeying the
same symmetries, the effective Einstein equations decouple into generalized
Friedmann and generalized Raychaudhuri equations. These were given in a form
insensitive to the particular embedding. Only the cosmological
\textquotedblleft constant\textquotedblright\ and the potential $U_{0}$
depend on the details of the embedding and bulk matter. While $\Lambda $ can
be found algebraically, the potential $U_{0}$ is determined by a first order
ordinary differential equation. An algorithm was given in the Appendix to
study cosmology on such generalized Friedmann branes.

With a definite choice of the bulk and embedding, the situation becomes even
simpler, the integration of the first order differential equation being
replaced by pure algebra. We have employed this advantage first in the case
of a Reissner-Nordstr\"{o}m-Anti de Sitter bulk, then for a charged
Vaidya-Anti de Sitter bulk. In both cases we have matched across the brane
inner/outer regions of the bulk space-times, finding the appropriate
generalized Friedmann and Raychaudhuri equations. The Raychaudhuri equation
acquires peculiar terms in the radiating case, which implies a non-standart
dependence of the density on the scale-factor. Our equations allow for
different mass and charge functions as well as bulk cosmological constants
on the two sides of the brane. However the junction does not allow for
different valiues of $k$ on the two sides, as $k$ is the curvature index in
the induced metric, required to be continuous.

The equations characterizing cosmological evolution both in the charged and
in the radiating case being given, the arena opens for imposing constraints
from experimental data on the non-symmetric character of the embedding as
well. In the low energy regime ($\rho \ll \lambda $), for example, $\Delta B$
contributes with the radiation-like term $\left( 3/2\widetilde{\kappa }%
^{2}\lambda ^{2}\right) \left( \Delta m\Delta \widetilde{\Lambda }%
/a^{4}\right) $ to the Friedmann equation and it is expected that
CMB-anisotropy data will constrain its magnitude, thus implicitly the non-$%
Z_{2}$-symmetric features of the bulk too.

While in the charged case only 3 cases should be considered (junctions of
inner/inner, inner/outer and outer/outer regions), in the radiating case the
direction of the radiation flow further diversifies the situation, leading
to a total of 10 cases to be discussed. Some of these will be ruled out by
energy conditions to be fulfilled on the brane. Further subtleties of the
radiating case include whether in the inner regions there are radiating
stellar objects, black holes or naked singularities (encountered in the 4d
Vaidya solution as well). Investigations into these issues are under way.

\section{Acknowledgments}

I am grateful to Roy Maartens for useful references, discussions and
comments on the manuscript and to Parampreet Sigh for a remark. This work
was supported by the Hungarian Scholarship Board and OTKA grant no. T034615.

\appendix

\section{Dynamics of perfect fluid Friedmann branes in arbitrary bulk. An
algorithm.}

Let us consider the physically interesting case of $d=4$. For a Friedmann
brane with perfect fluid, embedded in a non-$Z_{2}$ way into a bulk
containing some energy-momentum tensor $\widetilde{\Pi }_{cd}$, we present
here an algorithmic way of solving the relevant equations.

\textbf{Constraints on the bulk matter:}

\begin{align}
h_{ab}\Delta \left( g^{ac}{}n^{d}{}\widetilde{\Pi }_{cd}\right) & =0\ ,
\label{A1} \\
\Delta \left( u^{c}{}n^{d}{}\widetilde{\Pi }_{cd}\right) & =\dot{\rho}+3%
\frac{\dot{a}}{a}\left( \rho +p\right) \ .  \label{A2}
\end{align}%
Once solved by some choice for the matter fields in the bulk, we can pass to
the \textbf{constraints on the embedding:}

\begin{align}
\widetilde{\kappa }^{2}\overline{\left( g_{a}^{c}{}n^{d}\widetilde{\Pi }%
_{cd}\right) }& =\nabla _{c}\overline{K}_{a}^{c}-\nabla _{a}\overline{K}\ ,
\label{A3} \\
\Delta \left( n^{a}n^{b}\widetilde{\Pi }_{ab}\right) & =\left( p-\lambda
\right) \overline{K}+\left( \rho +p\right) u_{a}u_{b}\overline{K}^{ab}\ .
\label{A4}
\end{align}%
The mean value of the extrinsic curvature, once found, gives rise to the
trace and tracefree parts of the extrinsic curvature term $\overline{L}_{ab}$%
: 
\begin{align}
\overline{L}& =\overline{K}_{ab}\overline{K}^{ab}-\overline{K}^{2}\ ,
\label{A5} \\
\overline{L}_{ab}^{TF}& =\overline{K}_{ab}\overline{K}-\overline{K}_{ac}%
\overline{K}_{b}^{c}+\frac{\overline{L}}{4}g_{ab}\ ,  \label{A6}
\end{align}%
and therefore to the \textbf{cosmological constant}%
\begin{equation}
\Lambda =\frac{\kappa ^{2}\lambda }{2}-\frac{\overline{L}}{4}-\frac{%
\widetilde{\kappa }^{2}}{2}\overline{\left( n^{c}n^{d}\widetilde{\Pi }%
_{cd}\right) }\ .  \label{A7}
\end{equation}%
The first order differential equation determines \textbf{the unknown
potential} $U_{0}$: 
\begin{equation}
\kappa ^{2}\left( \frac{a_{0}}{a}\right) ^{4}\dot{U}_{0}+\dot{\Lambda}%
+\kappa ^{2}\left( 1+\frac{\rho }{\lambda }\right) \Delta \left(
u^{c}{}n^{d}{}\widetilde{\Pi }_{cd}\right) =0\ .  \label{A8}
\end{equation}%
Then $\Lambda $ and $U_{0}$ can be inserted into the generalized \textbf{%
Friedmann and Raychaudhuri equations}:%
\begin{align}
\frac{\dot{a}^{2}+k}{a^{2}}& =\frac{\Lambda }{3}+\kappa ^{2}\frac{\rho }{3}%
\left( 1+\frac{\rho }{2\lambda }\right) +\frac{\kappa ^{2}}{3}\left( \frac{%
a_{0}}{a}\right) ^{4}U_{0}\ ,  \label{A9} \\
\frac{\ddot{a}}{a}& =\frac{\Lambda }{3}-\frac{\kappa ^{2}}{6}\left[ \left(
1+2\frac{\rho }{\lambda }\right) \rho +3\left( 1+\frac{\rho }{\lambda }%
\right) p\right]  \notag \\
& -\frac{\kappa ^{2}}{3}\left( \frac{a_{0}}{a}\right) ^{4}U_{0}\ .
\label{A10}
\end{align}%
Up to this point, only the mean value of the extrinsic curvature has to be
solved for. (Still, its jump contributed implicitly to the functional form
of the generalized Friedmann and Raychaudhuri equations.)

Finally, in order to study the \textbf{off-brane evolution}, the extrinsic
curvature is needed on both sides on the brane%
\begin{equation}
2K_{ab}^{\pm }=2\overline{K}_{ab}\pm \Delta K_{ab}\ .  \label{A11}
\end{equation}%
Thus we have to determine its jump from the \textbf{Lanczos equation }%
\begin{equation}
3\Delta K_{ab}=-\widetilde{\kappa }^{2}\left[ \left( 2\rho +3p-\lambda
\right) u_{a}u_{b}+\left( \rho +\lambda \right) a^{2}h_{ab}\right] \ .
\label{A12}
\end{equation}%
The evolution in the off-brane direction of the brane gravitational
variables is determined by:%
\begin{eqnarray}
\mathcal{L}_{\mathbf{n}}g_{ab} &=&2K_{ab}^{\pm }\ ,  \label{A13} \\
\mathcal{L}_{\mathbf{n}}K_{ab}^{\pm } &=&K_{ac}^{\pm }K_{b}^{\pm c}-\frac{%
\widetilde{\kappa }^{2}}{3}\left( g_{a}^{c}{}g_{b}^{d}{}\widetilde{\Pi }%
_{cd}^{\pm }\right) ^{TF}  \notag \\
&&-\overline{\mathcal{E}}_{ab}\mp \frac{1}{2}\Delta \mathcal{E}_{ab}-\frac{%
g_{ab}}{4}E^{\pm }  \notag \\
&&+\left( \nabla _{b}\alpha _{a}-\alpha _{b}\alpha _{a}\right) ^{\pm }\ ,
\label{A14}
\end{eqnarray}%
with%
\begin{gather}
\overline{\mathcal{E}}_{ab}=\overline{L}_{ab}^{TF}+\frac{2\widetilde{\kappa }%
^{2}}{3}\overline{\left( g_{a}^{c}{}g_{b}^{d}{}\widetilde{\Pi }_{cd}\right)
^{TF}}  \notag \\
-\kappa ^{2}\left( \frac{a_{0}}{a}\right) ^{4}U_{0}\left( u_{a}u_{b}+\frac{%
a^{2}}{3}h_{ab}\right) \ ,  \label{A15} \\
\Delta \mathcal{E}_{ab}=\frac{2\widetilde{\kappa }^{2}}{3}\Delta \left(
g_{a}^{c}{}g_{b}^{d}{}\widetilde{\Pi }_{cd}\right) ^{TF}-\widetilde{\kappa }%
^{2}\Biggl[\left( \rho +p\right) \overline{K}u_{a}u_{b}  \notag \\
-2\left( \rho +p\right) u_{(a}\overline{K}_{b)}^{c}u_{c}-\frac{\rho
+3p-2\lambda }{3}\overline{K}_{ab}\Biggr]^{TF}\ ,  \label{A16} \\
E^{\pm }=\widetilde{\kappa }^{2}\left( n^{a}n^{b}\widetilde{\Pi }_{ab}^{\pm
}-\frac{1}{3}\widetilde{\Pi }^{\pm }\right) \ ,  \label{A17}
\end{gather}%
and $\alpha ^{b}=n^{c}\widetilde{\nabla }_{c}n^{b}$ an acceleration-like
quantity, the curvature of $n^{a}$, which can be freely specified.

\end{document}